\documentclass{ptephy_v1}

\usepackage[colorlinks=true,citecolor=blue,urlcolor=blue,linkcolor=blue]{hyperref}

\usepackage{graphicx}

\usepackage{amssymb}
\usepackage{amsmath}
\usepackage{breakurl,xspace,wasysym,multirow}

\newcommand{\nn}{\nonumber}
\newcommand{\beq}{\begin{equation}}
\newcommand{\eeq}{\end{equation}}
\newcommand{\beqa}{\begin{eqnarray}}
\newcommand{\eeqa}{\end{eqnarray}}
\newcommand{\TeV}{{\rm TeV}}
\newcommand{\GeV}{{\rm GeV}}

\def\lqcd{\Lambda_{\rm QCD}}
\def\d{{\rm d}}

\newcommand{\Bbar}{\,\overline{\!B}{}}
\newcommand{\Dbar}{\,\overline{\!D}{}}
\newcommand{\Kbar}{\,\overline{\!K}{}}
\def\B0bar{\Bbar{}^0}
\def\D0bar{\Dbar{}^0}
\def\K0bar{\Kbar{}^0}

\makeatletter
\g@addto@macro\bfseries{\boldmath}
\let\Hy@backout\@gobble
\makeatother

\begin{document}

\title{On magnitudes of some CKM matrix elements}


\author{Zoltan Ligeti}
\affil{Ernest Orlando Lawrence Berkeley National Laboratory, \\
University of California, Berkeley, CA 94720, USA \email{ligeti@berkeley.edu}}

\begin{abstract}%
This writeup follows the presentation at the Symposium, with emphasis on topics and ideas discussed there.  It is purposefully informal, not a review of the field, and neither does it include a complete list of references.
However, I hope that readers might find some comments useful or amusing, and may appreciate the challenges and reasons for excitement about recent progress and future opportunities in flavor physics.

\end{abstract}

\maketitle

\section{Introduction}

I was asked to talk at this symposium, celebrating the 50th anniversary of the
Kobayashi--Maskawa theory~\cite{Kobayashi:1973fv}, about the determinations of
the magnitudes of elements of the Cabibbo--Kobayashi--Maskawa (CKM) quark mixing
matrix~\cite{Kobayashi:1973fv, Cabibbo:1963yz}.  This talk
only covers (some of) those CKM elements that involve the third generation of quarks.

Given that much of the audience, as well as I, was not yet (or barely) alive when the paper that this Symposium celebrates was written, 
I would like to emphasize how stunning the timeline was.
After the discovery of $CP$ violation~\cite{Christenson:1964fg} in 1964, Wolfenstein's
paper on the superweak model~\cite{Wolfenstein:1964ks} was nearly instantaneous
(the PRL received dates are July~10 and August~31, 1964, respectively), whereas
PTEP received the Kobayashi--Maskawa manuscript~\cite{Kobayashi:1973fv} on September 1, 1972. 
I am old enough to remember people talking about excluding the superweak theory as a goal for Belle and BaBar, and young enough to never have taken that seriously.
Imagine, those 8 years between 1964 and 1972 were just as if ATLAS~\cite{TheATLAScollaboration:2015mdt} and CMS~\cite{CMS:2015dxe} had discovered a new particle near 750\,GeV in 2015, but all theory papers in the first 7 years had turned out to be wrong, and the correct model had only been proposed in 2023.

While this talk is about magnitudes of CKM elements, it is important to emphasize that 
testing the flavor sector of the standard model (SM), as a whole, is a lot more interesting than any single measurement.  It is the combination of many measurements that gives often the most interesting information, such as the constraints on beyond SM (BSM) contributions to flavor-changing neutral-current processes.
(Wolfenstein used to say that even though he invented the Wolfenstein parameters~\cite{Wolfenstein:1983yz}, he did not care what their values were, only whether their independent determinations gave consistent results.)
For the success of this program then, the interplay between experimental and theoretical developments, seeking to optimize theoretical cleanliness and experimental precision, has been crucial and also a lot of fun.

Since a substantial part of this talk concerns semileptonic decays, it may be amusing to note that semileptonic operators (to be precise, coefficients of operators composed of $l\bar l q\bar q$ fields) account for 1053 (i.e., 42\%) of the 2499 parameters
of the dimension-6 baryon and lepton number conserving terms in the 3-generation
SMEFT~\cite{Alonso:2013hga} (558 $CP$-even and 495 $CP$-odd terms).  And in the low
energy effective theory below the weak scale, there are 1944 semileptonic
parameters (i.e., 54\%) of the 3631 terms~\cite{Jenkins:2017jig} (1017 $CP$-even
and 927 $CP$-odd).\footnote{I thank Aneesh Manohar for help with adding (the correct) integers.}

\section{The Past}

To proceed from the Kobayashi--Maskawa proposal to building asymmetric $B$ factories, and pursuing their spectacularly rich physics program, required mixing angles and quark masses to have fortuitous values.  (Technological developments were also critical, of course.)  The combination of the value of $m_b$ (about half of the mass of the $\Upsilon$ resonance, discovered in 1977~\cite{E288:1977xhf}) and the smallness of $|V_{cb}|$ (discovered via the long lifetime of $b$-flavored hadrons in 1983~\cite{Fernandez:1983az, Lockyer:1983ev}) resulted in $b$ hadrons propagating long enough distances to be measured with detectors developed in the 1980s and 90s.  (If $|V_{cb}|$ was as large as the Cabibbo angle, $|V_{us}|$, then it would have been impossible to make time-dependent $CP$ violation measurements at Belle and BaBar in the early 2000s.)  Another critical ingredient was the much larger value of $m_t$ than anticipated in the 1980s, enabling the ARGUS discovery of $B^0$\,--\,$\B0bar$ oscillation~\cite{ARGUS:1987xtv}.
The measurement of the ratio of $B^0$ decays after mixing vs.\ unmixed decays, $r = 0.21 \pm 0.08$, implied that $m_t$ was much greater than the direct bound in 1987, $m_t > 23\,\GeV$.
The comparable time scales of oscillation and decay, $\Delta m/\Gamma = 0.77$, and constraining $\Delta\Gamma \ll \Gamma$ for $B^0$ mixing were also important.  Finally, CLEO~\cite{CLEO:1989gsk} and ARGUS~\cite{ARGUS:1989eae} established a nonzero value for $|V_{ub}|$ in 1989; if it were zero, the CKM matrix could not contain a physical $CP$ violating phase, as the Jarskog invariant is proportional to $|V_{ub}|$.  These were all crucial to make (asymmetric) $B$ factories exciting and plausible to pursue.

\begin{figure}[tb]
\centering{\includegraphics[width=.5\textwidth]{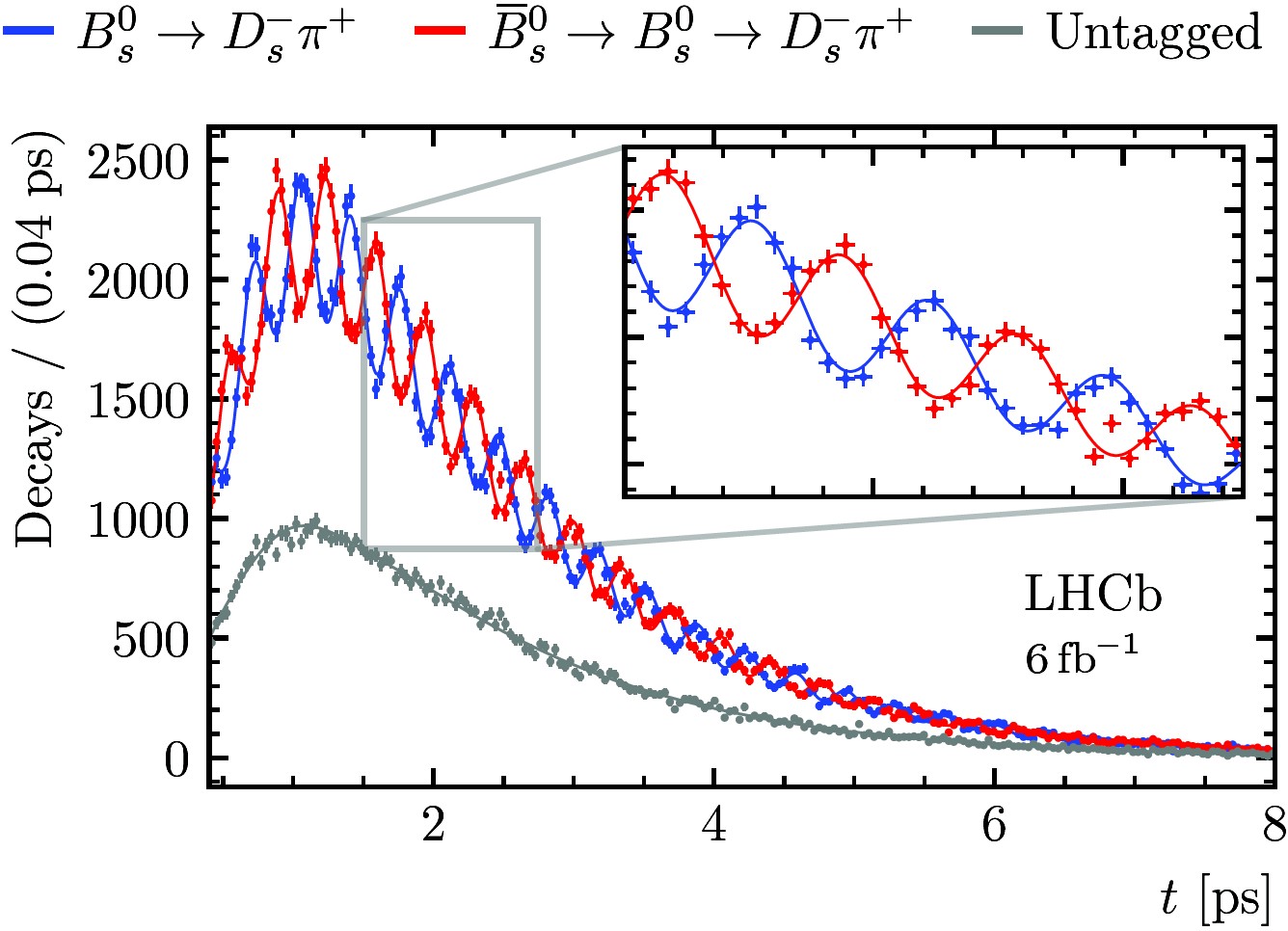}}
\vspace*{-8pt}
\caption{Time-dependence of ``right-sign'' (blue) and ``wrong-sign'' (red) $B_s$ decays~\cite{LHCb:2021moh}.}
\label{fig:Bsmix}
\end{figure}

The mixing of neutral mesons plays a special role in determining CKM matrix elements and constraining BSM scenarios.  This goes back to the successful prediction of $m_c$ from $K^0$\,--\,$\K0bar$ mixing~\cite{Vainshtein:1973md, Gaillard:1974hs}.  Since the 1970s, $\Delta m_K$ and $CP$ violation in the kaon sector have provided some of the strongest constraints on BSM scenarios, because flavor-changing neutral currents (FCNCs) are most strongly suppressed between the 1st and 2nd generation of quarks.  In the $B^0$ and $B_s^0$ systems the large top mass implies that the GIM mechanism~\cite{Glashow:1970gm} does not yield strong suppressions, and FCNC decay rates and $CP$ violating observables can be much larger than in $K^0$ or $D^0$ meson processes.  After the initial measurement of $B_s$ oscillation by CDF~\cite{CDF:2006imy}, by now $\Delta m_{B_s} = (17.7656 \pm 0.0057)\,{\rm ps}^{-1}$~\cite{LHCb:2021moh} is one of the most precise CKM measurements, yielding a $1.6\times 10^{-4}$ relative uncertainty for $|V_{tb}V_{ts}|\, f_{B_s}^2 B_{B_s}$.  The LHCb Collaboration's textbook measurement of the time dependence of unmixed (``right-sign'') and mixed (``wrong-sign``) $B_s$ decays is shown in Figure~\ref{fig:Bsmix}.  If the uncertainty of $f_{B_s}^2 B_{B_s}$ from lattice QCD did not dominate, the uncertainty of $|V_{tb}V_{ts}|$ would compete with that of $|V_{ud}|$ (the latter may actually be underestimated~\cite{Seng}).

The evaluation of FCNC $B$ decay amplitudes at the (multi-)loop level also started around 1986--87, including two-loop calculations of $b\to s\gamma$~\cite{Grinstein:1987vj, Bertolini:1986th, Deshpande:1987nr}, which is important both for its sensitivity to BSM physics, and, in the SM, to $|V_{tb}V_{ts}|$.  Remarkably, for $m_t \simeq 170\,\GeV$ (unknown at that time), the prediction 
$\Gamma(B\to X_s\gamma) / \Gamma(B\to X_c e\bar\nu) \approx 0.003$~\cite{Grinstein:1987vj} is very close to the current central value~\cite{Misiak:2015xwa}.  In the presence of unavoidable experimental phase space cuts, the $b$-quark distribution function in the $B$ meson~\cite{Bigi:1993ex, Neubert:1993um} becomes important to describe the photon spectrum, which is largely BSM physics independent~\cite{Kapustin:1995nr}, can be extracted from the measured decay distributions~\cite{Bernlochner:2020jlt}, and plays a crucial role in the determination of $|V_{ub}|$ from inclusive decays.

The development of heavy quark symmetry~\cite{Isgur:1989vq, Isgur:1990yhj} and heavy quark effective theory (HQET)~\cite{Georgi:1990um} also started around 1990, as well as putting the calculation of inclusive $B$ decay rates on rigorous foundations~\cite{Chay:1990da}. These have played crucial roles since then, for the determinations of $|V_{cb}|$ and $|V_{ub}|$, which are essential to constrain BSM physics by comparing measurements dominated by tree-level processes in the SM with loop-mediated FCNC measurements.  In particular, the uncertainty of $|V_{cb}|$ is a large part of the uncertainty of the SM predictions for $\epsilon_K$, ${\cal B}(K\to \pi\nu\bar\nu)$, ${\cal B}(B_{d,s}\to \ell^+\ell^-)$, etc.  The measurement of $|V_{ub}|$ is especially important in constraining BSM physics, because it constitutes the dominant uncertainty in the side of the unitarity triangle (a graphical representation of the $V_{ud}\, V_{ub}^* + V_{cd}\, V_{cb}^* + V_{td}\, V_{tb}^* = 0 $ relation; see Figure~\ref{fig:UTfits}) opposite to the currently best measured angle, $\beta (\equiv \phi_1)$.  Moreover, the same theoretical tools can be used to analyze inclusive and exclusive FCNC decays, with complementary sensitivity to BSM physics, such as those mediated by $b\to s\gamma$, $b\to s\ell^+\ell^-$, and $b\to s\nu \bar\nu$ transitions.
(A recent illustration of how such combinations work is the semi-inclusive analysis of $B\to X_s \ell^+\ell^-$ in the high-$q^2$ region, where the best constraints are derived~\cite{Isidori:2023unk} by comparing the data with a similar Belle measurement of $B\to X_u\ell\bar\nu$~\cite{Belle:2021ymg}, using the method of Ref.~\cite{Ligeti:2007sn}.)


Since the development of HQET, the determination of $|V_{cb}|$ from exclusive $B\to D^{(*)}\ell\bar\nu$ decays has relied on measuring the $q^2=(p_\ell+p_\nu)^2$ distribution near the so-called zero recoil point, $w=1$  [$w= (m_B^2 +m_{D^{(*)}}^2-q^2)/(2m_B m_{D^{(*)}})$], where the $D^{(*)}$ is at rest in the restframe of the decaying $B$ meson.
The rates can be schematically written as
\begin{equation}\label{rates}
{\d\Gamma(B\to D^{(*)} \ell\bar\nu)\over \d w} 
  = (\mbox{calculable terms})\, |V_{cb}|^2
\begin{cases}
(w^2-1)^{1/2}\, {\cal F}_*^2(w), & \mbox{for } B\to D^*, \\
(w^2-1)^{3/2}\, {\cal F}^2(w), & \mbox{for } B\to D\,. 
\end{cases}
\end{equation}
Both ${\cal F}(w)$ and ${\cal F}_*(w)$ are equal to the Isgur-Wise function in
the $m_{b,c} \gg \lqcd$ limit, and ${\cal F}_{(*)}(1) = 1$ is the basis for a
model-independent determination of $|V_{cb}|$.  There are calculable 
corrections in powers of $\alpha_s(m_{c,b})$, as well as terms suppressed by
$\lqcd/m_{c,b}$, which can only be parametrized, and that is where hadronic
uncertainties enter.  Schematically,
\beqa\label{F1}
{\cal F}_*(1) &=& 1_{\mbox{\footnotesize (Isgur-Wise)}} + c_A(\alpha_s) 
  + {0_{\mbox{\footnotesize (Luke)}}\over m_{c,b}} 
  + {\cal O}(\lqcd^2/m_{c,b}^2) \,, \nn\\
{\cal F}(1) &=& 1_{\mbox{\footnotesize (Isgur-Wise)}} + c_V(\alpha_s) 
  + {\cal O}(\lqcd/m_{c,b}) \,.
\eeqa
The absence of the ${\cal O}(\lqcd/m_{c,b})$ term for $B\to D^*\ell\bar\nu$ at
zero recoil is a consequence of Luke's theorem~\cite{Luke:1990eg}.  Calculating
corrections to the heavy quark limit in these decays is a vast subject.  
To achieve percent level precision, lattice QCD appears to be the only tool~\cite{El-Khadra}.  FLAG~\cite{FlavourLatticeAveragingGroupFLAG:2021npn} quotes ${\cal F}_*(1) = 0.904 \pm 0.012$ and ${\cal F}(1) = 1.054 \pm 0.009$, while the experimental data are $|V_{cb}\, {\cal F}_*(1)| = (34.77 \pm 0.36)\times 10^{-3}$ and $|V_{cb}\, {\cal F}(1)| = (41.26 \pm 0.97)\times 10^{-3}$~\cite{HFLAV:2022esi}.  An important ingredient of the analysis which only received full attention in recent years, after the publication of unfolded measurements~\cite{Belle:2017rcc, Belle:2018ezy}, is related to the functional form of fitting the data and extrapolating the rate to $w=1$ (where phase space vanishes).  In the experimental determinations of $|V_{cb}|$ before 2017, the model-independent Boyd-Grinstein-Lebed (BGL) parametrization~\cite{Boyd:1995cf, Boyd:1995sq, Boyd:1997kz} was supplemented by QCD sum rule results~\cite{Neubert:1992wq, Neubert:1992pn, Ligeti:1993hw} for the subleading ${\cal O}(\lqcd/m_{c,b})$ Isgur-Wise functions, to arrive at a fit prescription with fewer parameters~\cite{Caprini:1997mu} (with hard to quantify model dependence), which all prior BaBar and Belle analyses used.  Recently two approaches were proposed to incorporate some ${\cal O}(\lqcd^2/m_c^2)$ corrections~\cite{Bordone:2019vic, Bernlochner:2022ywh}.  The question how to fix the number of BGL expansions parameters that are fit from a measurement of $B\to D^*\ell\bar\nu$ distributions with a given statistics, is subject to debate~\cite{Bernlochner:2019ldg, Gambino:2019sif, Simons:2023wfg}, and different choices in recent experimental analyses~\cite{Belle:2023bwv, Belle-II:2023okj} impact the extracted value of $|V_{cb}|$ significantly.  Thus, in future measurements a detailed validation of the choices made with toy Monte Carlo studies seems essential.  There are additional open questions related to (i) moderate tensions between HQET predictions, lattice QCD results and form factor measurements; (ii) the role of the $D^{**}$ states, etc.~\cite{ZLoct23}.

For the determination of $|V_{cb}|$ from inclusive $B\to X_c\ell\bar\nu$ decay, instead of identifying all particles, final state hadrons that can be produced by the strong interaction are summed over, subject to constraints determined by short-distance physics, e.g., the energy of the charged lepton.  Although hadronization is nonperturbative, it occurs on much longer distance (and time) scales than the underlying
weak decay, and therefore one can use an operator product expansion (OPE) to calculate the decay rate in a systematic expansion.  The leading order corresponds to the free $b$-quark decay rate, and perturbative and nonperturbative corrections can systematically be accounted for.  The first few dominant nonperturbative parameters can be extracted by fitting differential decay distributions in the same $B\to X_c\ell\bar\nu$ decays, an approach that has been pursued for over twenty years now (mainly using fits in the so-called $1S$~\cite{Bauer:2004ve, Bauer:2002sh} and kinetic~\cite{Alberti:2014yda, Gambino:2004qm} mass schemes).
Recently 3-loop $\alpha_s^3$ corrections were calculated~\cite{Fael:2020njb, Fael:2020tow, Bordone:2021oof}, $\alpha_s$ corrections are known up to ${\cal O}(\lqcd^3/m^3)$~\cite{Mannel:2021zzr}, and lattice QCD calculations of inclusive rates are being developed~\cite{Barone:2023tbl}.

Given that determinations of $|V_{cb}|$ from exclusive and inclusive decays have been in persistent tension (at an inconclusive, couple of sigma level), it is possible that a detailed understanding of the composition of the inclusive rate in terms of the sum over exclusive modes will be needed, in order to understand the resolution of this tension (whether on the theoretical or experimental side, or a combination).  The upcoming much larger data sets should allow this to be explored and hopefully resolved.

Determinations of $|V_{cb}|$, that are currently less precise, can also be obtained from the measurement of $B_s \to D_s^{(*)} \mu\bar\nu$ decay~\cite{LHCb:2020cyw}.
In addition, semileptonic decays to $\tau$ leptons measured in
$B\to D^{(*)}\tau\bar\nu$ and related modes are also sensitive to $|V_{cb}|$.  
The most precise data involving $\tau$ leptons are the $|V_{cb}|$-independent ratios, ${\cal B}(B\to D^{(*)}\tau\bar\nu) / {\cal B}(B\to D^{(*)}\ell\bar\nu)$
measured by BaBar, Belle, and LHCb for the last ten years.
If the current, approximately $3\,\sigma$~\cite{HFLAV:2022esi}, hint of lepton 
non-universality prevails, the determination of $|V_{cb}|$ becomes more complicated, since one would need to know the size of the BSM contributions in the $e$ and $\mu$ modes.
It is due to heavy quark symmetry that the SM predictions for these ratios of rates is under good theoretical control~\cite{Bernlochner:2017jka, Bernlochner:2022ywh}, even independent of lattice QCD.

\section{The Present}

\begin{figure}[b]
\centering\includegraphics[width=.7\textwidth, clip, bb=40 0 480 335]{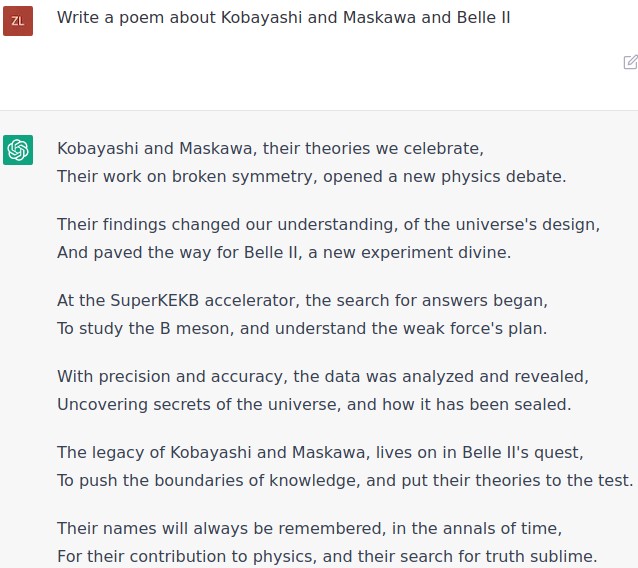}
\caption{ChatGPT: write a poem about Kobayashi, Maskawa, and Belle~II~(1/29/2023).}
\label{fig:poem}
\end{figure}

There was little to say about the present at the time of this Symposium, in February 2023, since neither LHC nor Belle was taking data, during LHC's LS2 and the Belle~II upgrade.  However, ChatGPT was fairly new, and I was amused by the poem it composed about Kobayashi, Maskawa, and Belle~II, shown in Fig~\ref{fig:poem}.  (I said that this was the first and last time I'd use ChatGPT in a talk, which I hope to continue to obey.)

\begin{figure}[tb]
\centering\includegraphics[width=.47\textwidth]{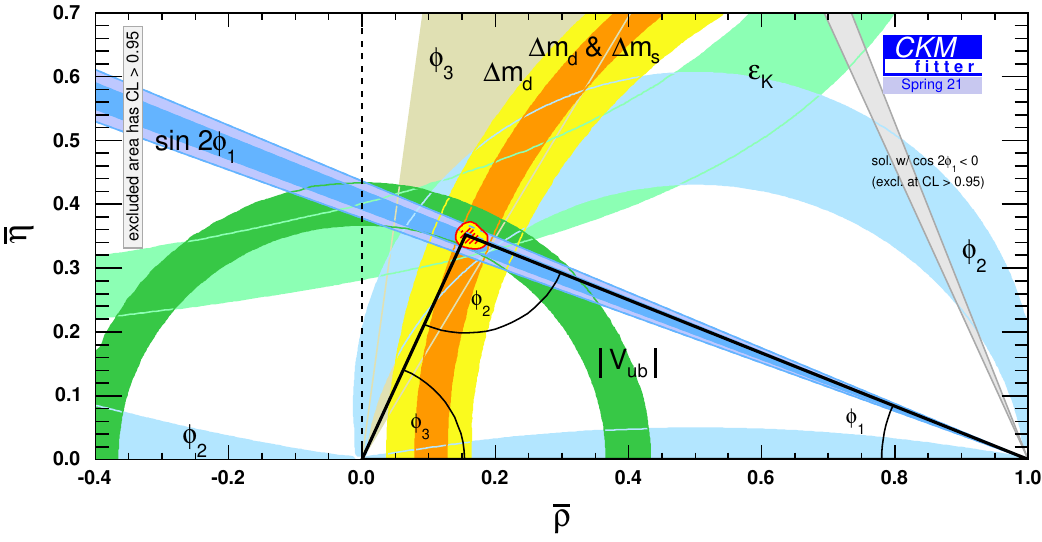} \hspace*{.25cm}
\includegraphics[width=.47\textwidth]{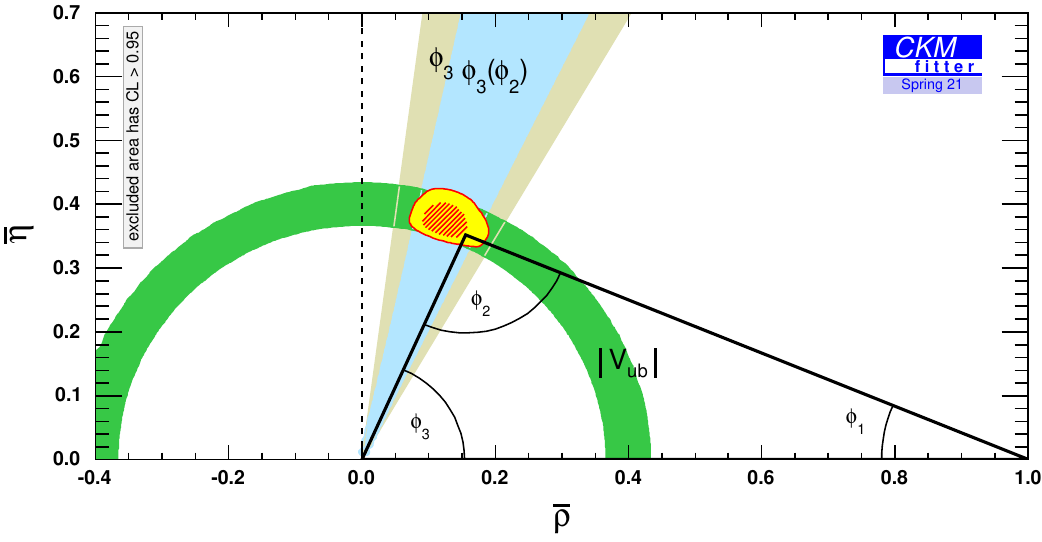}
\vspace*{-8pt}
\caption{Main constraints on the unitarity triangle (left), and their subset dominated by tree-level SM processes (right).  The colored regions show 95\% CL.  (From
Ref.~\cite{Charles:2004jd}.)}
\label{fig:UTfits}
\end{figure}

Possibly the most concise illustration of the achievements of the last two decades is what used to be the ``money plot'' for BaBar and Belle, summarizing the main constraints on the unitarity triangle, shown in Figure~\ref{fig:UTfits} (left plot).  These measurements of primarily $B_{u,d}$ decays (some of them further refined by LHCb) established the Kobayashi--Maskawa mechanism as the dominant source of $CP$ violation in flavor changing processes in the quark sector.
Although this SM CKM fit shows impressive and nontrivial consistency, the implications of the agreement among different measurements are often overstated.  The determination of CKM parameters only using dominantly tree-level decays yields the plot on the right in Figure~\ref{fig:UTfits}.  The constraints shown in the left plot in Figure~\ref{fig:UTfits} but not on the right, all involve neutral meson mixing.  Allowing BSM contributions, FCNC processes may be altered significantly, and additional parameters related to $CP$ and flavor violation arise, so the fits become less constraining.  The conclusion of such analyses, some of which will be discussed below, is that ${\cal O}(20\%)$ BSM contributions to most FCNC processes are still allowed.  As a result of LHCb measurements in Run 1 and 2, the constraints in the $B_s$ sector have become comparably strong to constraints in the $B_d$ sector.

\section{The Future}

In the next one to two decades, the LHC experiments~\cite{Bediaga:2018lhg, Cerri:2018ypt}, Belle~II~\cite{Belle-II:2018jsg}, BES\,III~\cite{BESIII:2022mxl} and their planned and possible upgrades will increase $b$-, $c$-, and $\tau$-decay data sets by almost two orders of magnitude.  
If the FCC (future circular collider) is built, it will have an extremely rich flavor physics program~\cite{FCC:2018byv, Grossman:2021xfq}, studying $10^{12}$ $b$-flavored hadrons from the decay of $Z$ bosons, in a clean experimental environment.  An intriguing aspect of the FCC-$ee$ flavor physics program is that in its $W^+W^-$ phase, the decay of $10^8$ $W$-s will allow a determination of $|V_{cb}|$ with about 0.3\% precision~\cite{Azzurri}, independent of $|V_{cb}|$ measurements in $B$ decays.


The measurement of $|V_{ub}|$ from inclusive semileptonic decays belongs in my mind largely to the future, because the existing determinations all include some unaccounted model dependence, which we know how to eliminate.  (The determination of $|V_{ub}|$ from exclusive decays seems fully in the domain of lattice QCD, at the relevant precision~\cite{El-Khadra}.)  
Similar to the case of $B\to X_s\gamma$ discussed above, also in the analysis of inclusive $B\to X_u\ell\bar\nu$ decays, significant experimental cuts on the phase space are required to suppress the much larger $B\to X_c\ell\bar\nu$ backgrounds.  As a result, instead of local HQET matrix elements, the nonperturbative quantities which determine the rate are the $b$-quark distribution functions in the $B$ meson~\cite{Bigi:1993ex, Neubert:1993um} (except for the $q^2 > (m_B-m_D)^2$ cut~\cite{Bauer:2000xf}).  At leading order in $\lqcd/m_b$ there is only one such function, which is identical for $B\to X_s\gamma$ and $B\to X_u\ell\bar\nu$, whereas at subleading order, several different distribution functions occur.  For $B\to X_u\ell\bar\nu$, the consistent treatment of the shape function region (where the hadronic final state is ``jetty'') and the local OPE region is more complicated than in the case of $B\to X_s\gamma$.

\begin{figure}[tb]
\centering\includegraphics*[width=.4\textwidth]{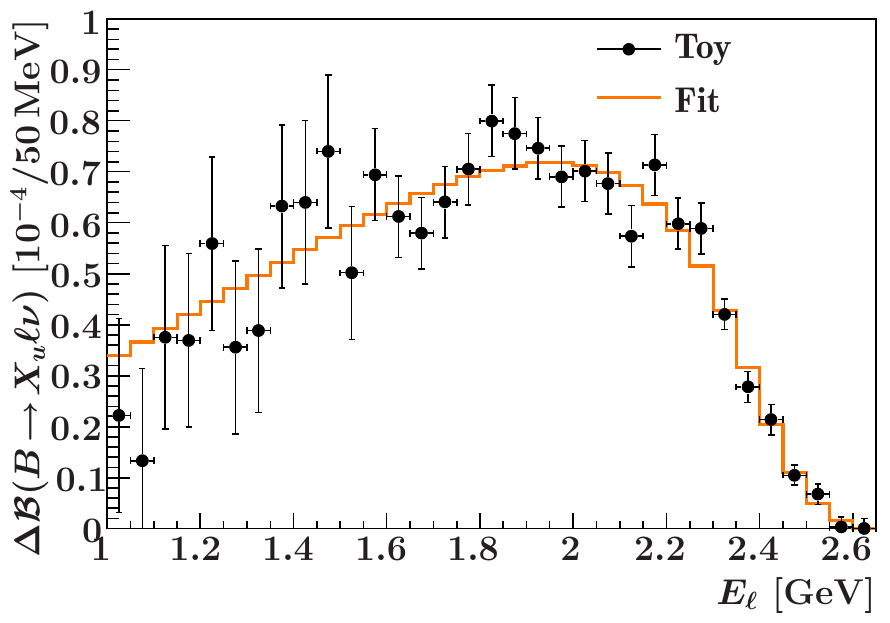} \hfil
\includegraphics*[width=.4\textwidth]{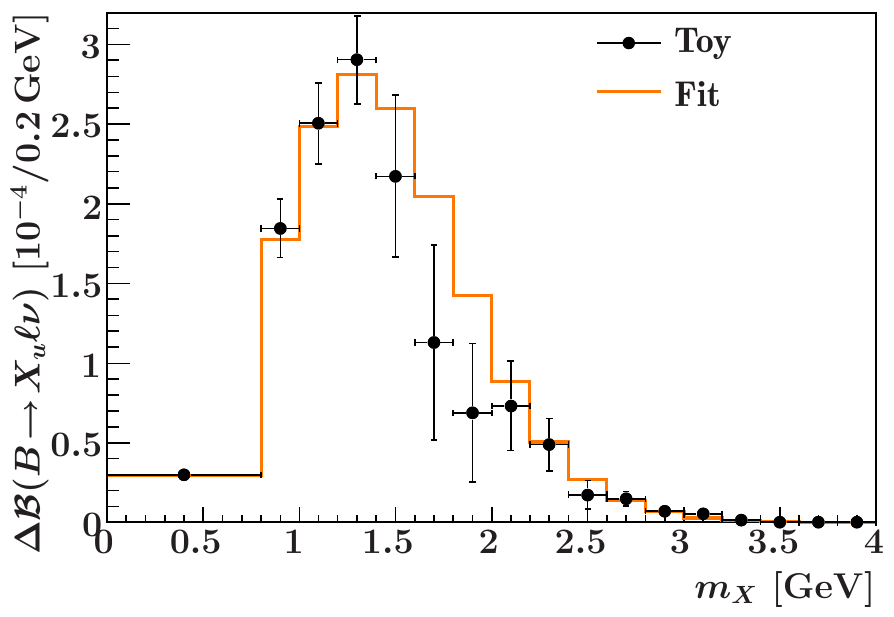}
\vspace*{-8pt}
\caption{A (not state-of-the-art) Monte Carlo study with SIMBA, fitting the charged lepton (left) and hadronic invariant mass (right) spectra, with statistics corresponding to 75/ab~\cite{Kerstin}.}
\label{fig:VubToys}
\end{figure}

Just like the determination of $|V_{cb}|$ from inclusive decays relies on extracting HQET matrix elements and $|V_{cb}|$ itself from the measurements of $B\to X_c\ell\bar\nu$ absolute rates and distributions, the determination of $|V_{ub}|$ from inclusive decays should utilize a similar strategy, fitting $B\to X_u\ell\bar\nu$ rates and distributions (as well as the $B\to X_s\gamma$ spectrum), to $b$-quark distribution functions and $|V_{ub}|$.  To develop such a self-consistent theoretical approach and fitting tool has been the goal of SIMBA~\cite{Bernlochner:2020jlt}.  Figure~\ref{fig:VubToys} shows a toy Monte Carlo study (as an illustration, not using the state of the art theory yet), fitting the $B\to X_u\ell\bar\nu$ charged lepton (left) and hadronic invariant mass (right) spectra, with statistics corresponding to 75/ab~\cite{Kerstin}.  The sensitivity is actually underestimated, as it assumed uncertainties and correlations similar to a BaBar full hadronic reconstruction analysis, compared to which the Belle~II hadronic tagging efficiency is already better.  This suggests that depending on the ability to constrain (linear combinations of) subleading shape functions, an uncertainty on $|V_{ub}|$ at the few percent level may be achievable at Belle~II.

As a slight aside, the measurements of $B_{d,s}\to \mu^+\mu^-$ are usually discussed for their BSM sensitivity, for good reasons.  While ${\cal B}(B_s\to \mu^+\mu^-) = (2.95 \pm 0.41)\times 10^{-9}$~\cite{HFLAV:2022esi} has been measured by three LHC collaborations, the decay $B_d\to \mu^+\mu^-$, which has a branching ratio around $10^{-10}$ in the SM, has not been observed.  Its uncertainty is not expected to reach 10\% even by the end of HL-LHC data taking.  It is an example of a channel, whose interpretation will not become theory limited in any planned experiment, and greater statistics would always teach us about short-distance physics.  In terms of an operator analysis, its mass-scale sensitivity is comparable to $K\to \pi\nu\bar\nu$.  Amusingly, the theoretically cleanest determination of $|V_{ub}|$ I know of, in principle, setting experimental realities aside, could come from the ratio ${\cal B}(B^-\to \ell\bar\nu) / {\cal B}(B^0\to \mu^+\mu^-)$ relying only on perturbative QCD calculations and isospin.

\subsection{Beyond SM sensitivity in $B$\,--\,$\Bbar$ mixing}

\begin{figure}[tb]
\centerline{
\includegraphics[width=.45\textwidth]{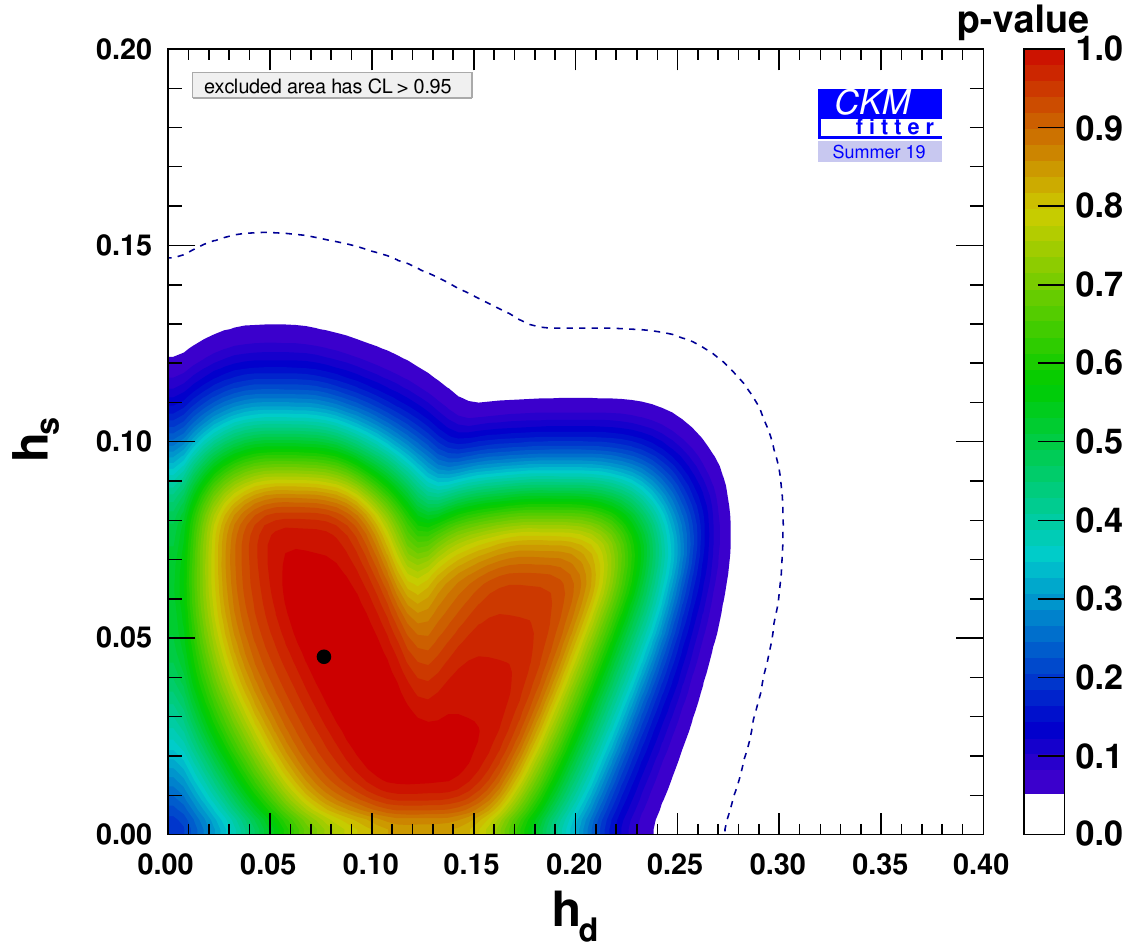}\hfil
\includegraphics[width=.41\textwidth]{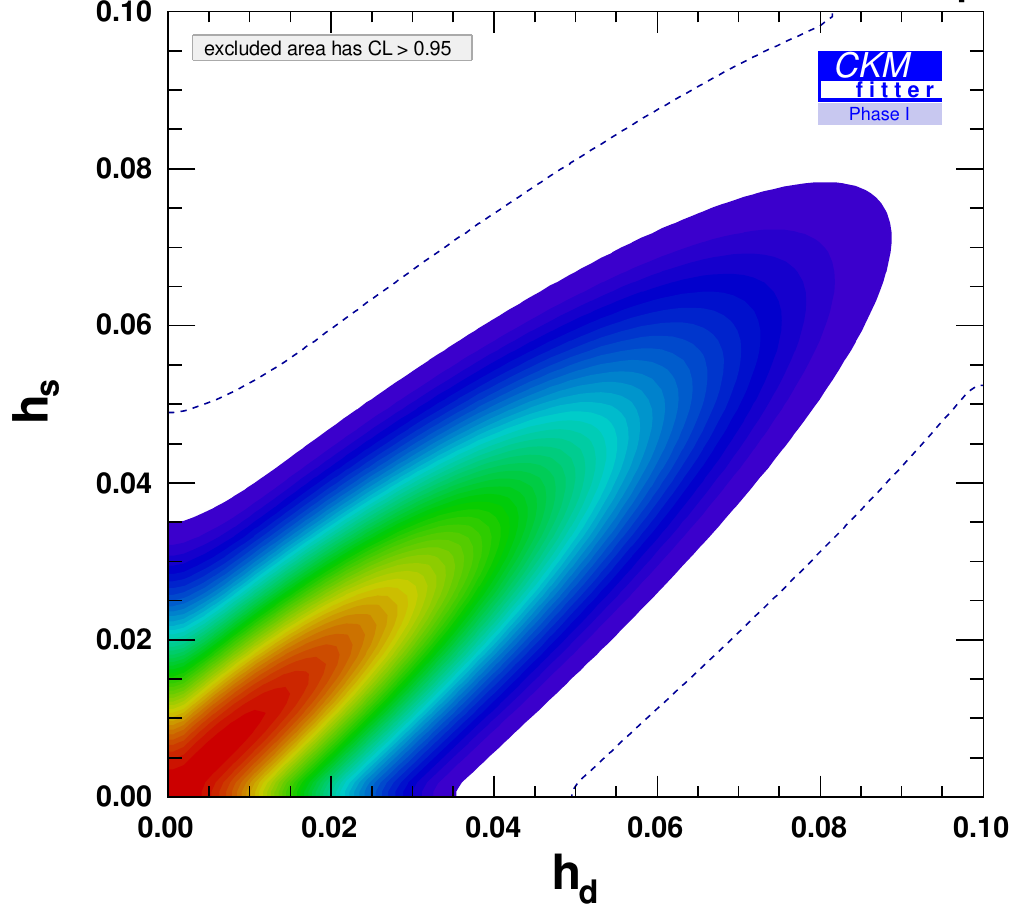}}
\caption{Constraints on the $h_d - h_s$ at present (left plot) and expected future sensitivity assuming 50/ab LHCb data, and 50/ab Belle~II data (right plot).
The colors indicate the confidence levels as shown, while the dashed lines show $3\sigma$ limits.  (From Ref.~\cite{Charles:2020dfl}.)}
\label{fig:NPmix}
\end{figure}

It has long been known that the mixing of neutral mesons is particularly sensitive to BSM physics, and probe some of the highest scales (the smallness of $\Delta m_K /m_K \approx 7 \times 10^{-15}$ is known since the 1960s).  
In a large classes of models, the dominant effect of BSM physics is to modify the mixing amplitudes of neutral mesons, and tree-level decays are not affected.  In this case, the $3\times 3$ CKM matrix remains unitary, and the BSM effects can be parametrized by two real parameters for each neutral meson system.
The mixing of $B_q^0$ mesons (where $q=d,s$) are simplest to analyze, as they are dominated by short-distance physics.  Writing the mixing amplitude as $M_{12}^q = M_{12\, \rm (SM)}^q\, \big(1 + h_q\, e^{2i\sigma_q}\big)$, one has to redo the CKM fit including the effects of $h$ and $\sigma$ on the observables.
The resulting constraints on $h_d$ and $h_s$, the magnitudes of the BSM contributions relative to the SM in $B_d$ and $B_s$ mixing, respectively, are shown in the left plot in Figure~\ref{fig:NPmix}, assuming independent BSM contributions.
This shows that order $(10-20)\%$ corrections to $M_{12}$ are still allowed. 
(Evidence for $h_q \neq 0$ would rule out the SM, and the black dot indicating the best-fit point in the left plot shows a slight pull.)  Similar conclusions apply to other neutral meson mixings, as well as many other $\Delta F=1$ FCNC decays, such as $B\to X\gamma$, $B\to X\ell^+\ell^-$, $B_{d,s}\to \ell^+\ell^-$, $K\to \pi\nu\bar\nu$, etc.
The same analysis, for a future time, using sensitivity projections for 50/ab LHCb data and 50/ab Belle~II data~\cite{Charles:2020dfl} (referred there as ``Phase~I'') are shown in the right plot in Figure~\ref{fig:NPmix}.  With even higher statistics expected with the LHCb Upgrade~II~\cite{LHCb:2022ine} and a possible upgrade of Belle~II~\cite{Forti:2022mti} (and especially FCC-$ee$), improving the determination of $|V_{cb}|$ beyond current projections becomes crucial for this type of analysis to continue to improve its sensitivity~\cite{Charles:2020dfl}.  If new developments result in theory uncertainties not becoming a limiting factor for this type of analysis, then the statistical sensitivity could reach~\cite{Charles:2020dfl} $h_d \lesssim 0.02$ and $h_s \lesssim 0.01$ at 95\%~CL, with the full LHCb Upgrade~II and the possible Belle~II upgrade datasets.

The correspondence between the constraints on $h_{d,s}$ and the scale of BSM physics depends on the operator(s) that contribute.  As the simplest example, if new contributions add a term $({C_{q}^2}/{\Lambda^2})\, (\bar b_{L}\gamma^{\mu}q_{L})^2$ to the same operator that describes $B$ mixing in the SM, then one finds
\beq
h_q \simeq \frac{|C_{q}|^2}{|V_{tb}^*\, V_{tq}|^2} 
  \left(\frac{4.5\, \TeV}{\Lambda}\right)^2\,.
\eeq
It is then straightforward to translate the constraints in Figure~\ref{fig:NPmix} to the scale of BSM physics probed, and the summary of expected sensitivities are shown in Table~\ref{tab:scales}, for different structures.   
These sensitivities, even with SM-like loop- and CKM-suppressed coefficients, are comparable to the scales probed by the LHC.
The factor of 1.5\,--\,2 increase in mass scale sensitivity at Phase~I compared to the present is greater than the improvement for many other searches during the HL-LHC.

\begin{table*}[bt]
\tabcolsep 4pt
\resizebox{\textwidth}{!}{
\begin{tabular}{c|c||c|c|c|c}
\hline\hline
\multirow{2}{*}{Couplings}  &  ~BSM loop~  &  
  \multicolumn{2}{c|}{~Sensitivity for Summer~2019 [TeV]~} &  \multicolumn{2}{c}{~Phase~I Sensitivity [TeV]~} \\
&  order  &  ~$B_d$ mixing~  &  $B_s$ mixing   &  ~$B_d$ mixing~  &  $B_s$ mixing  \\
\hline
$|C_{ij}| = |V_{ti}V_{tj}^*|$ & tree level 
  & 9 & 13 & 17 & 18 \\
(CKM-like)  &  one loop  
  & 0.7 & 1.0 & 1.3 & 1.4 \\
\hline
$|C_{ij}| = 1$  &  tree level 
  & $1\times 10^3$ & $3\times 10^2$ & $2\times 10^3$ & $4\times 10^2$ \\
(no hierarchy)  &  one loop 
  & 80 & 20 & $2\times 10^2$ & 30 \\
\hline\hline
\end{tabular}}
\caption{The scale of the $({C_{q}^2}/{\Lambda^2})\, (\bar b_{L}\gamma^{\mu}q_{L})^2$ operator probed (in TeV, at 95\%~CL) by $B_d$ and $B_s$ mixings at present and at ``Phase~I'' (described in the text)~\cite{Charles:2020dfl}.  The impact of SM-like hierarchy of couplings and/or loop suppression is shown.}
\label{tab:scales}
\end{table*}

\section{Final Remarks}

Pursuing ever more precise tests of the flavor sector remains exciting and well motivated.  If there is BSM physics below the few tens of TeV scale, which we hope to probe directly in current and planned experiments, they cannot have a generic flavor structure, as that is already ruled out by orders of magnitudes.  There is thus an unavoidable complementarity between direct searches of BSM physics, and learning about their signatures (or lack of signatures) in precision experiments.  In a large class of scenarios one may expect deviations from the SM close to the current experimental precision, so there is significant discovery potential any time experimental sensitivities improve substantially.

As in the past, theory will be important for the interpretation of the measurements and for maximizing their sensitivity to BSM physics, and experimental results will be essential inputs of theory considerations, and triggers for new developments.
In particular, the reduction of the uncertainties of $|V_{cb}|$ and $|V_{ub}|$ are crucial.  In each case, with new measurements, the tension between inclusive and exclusive measurements will hopefully shrink, and possibly yet unknown subtleties understood, and the source of current tensions identified.  They may lie either on the theory side or on the experimental side, or a combination.  In the case of the determination of $|V_{ub}|$ from inclusive decays, we know how qualitatively better measurements can be done, than those implemented so far.  And the determinations from exclusive semileptonic decays will improve with statistics and better lattice QCD calculations.  Both for $|V_{cb}|$ and $|V_{ub}|$, measurements from two-body leptonic decays will also become available, and will become competitive in precision, as statistics grows.

These experimental improvements need much larger data sets than available so far, and there is a suite of measurements (which have been identified, and will likely become further enlarged by future theory developments), in which theoretical uncertainties will not limit probing short-distance physics.  
However, I do not think anyone has seriously explored what the largest $B$ decay data sets may be, which would improve sensitivity to short-distance physics.\footnote{I vividly remember Sanda-san at the Workshops on Higher Luminosity $B$ Factory at Izu in 2003~\cite{Sanda}, saying (as a joke? or not as a joke?) that the question was not whether the design luminosity of the next generation $B$ factory should be $10^{35}$ or $10^{36}$, but rather if it should be $10^{53}$ or $10^{63}$.}  My understanding is that the sensitivity to possible signatures of BSM physics would continue to improve with data sets well beyond all upgrades contemplated at LHCb and Belle~II, and even at the Tera-$Z$ run of the FCC.

\section*{Acknowledgments}

I thank the organizers for the invitation to this very enjoyable Symposium, and for their patience regarding the completion of this writeup.
I thank Shoji Hashimoto and Aneesh Manohar for helpful conversations.
This work was supported in part by the Office of High Energy Physics of the U.S.\ Department of Energy under contract DE-AC02-05CH11231.

\newpage
\bibliographystyle{utphys}

\bibliography{refs}

\providecommand{\href}[2]{#2}\begingroup\raggedright\begin{thebibliography}{10}

\bibitem{Kobayashi:1973fv}
M.~Kobayashi and T.~Maskawa, ``{CP Violation in the Renormalizable Theory of
  Weak Interaction},'' \href{http://dx.doi.org/10.1143/PTP.49.652}{{\em Prog.
  Theor. Phys.} {\bfseries 49} (1973) 652--657}.

\bibitem{Cabibbo:1963yz}
N.~Cabibbo, ``{Unitary Symmetry and Leptonic Decays},''
  \href{http://dx.doi.org/10.1103/PhysRevLett.10.531}{{\em Phys. Rev. Lett.}
  {\bfseries 10} (1963) 531--533}.

\bibitem{Christenson:1964fg}
J.~H. Christenson, J.~W. Cronin, V.~L. Fitch, and R.~Turlay, ``{Evidence for
  the $2\pi$ Decay of the $K_2^0$ Meson},''
  \href{http://dx.doi.org/10.1103/PhysRevLett.13.138}{{\em Phys. Rev. Lett.}
  {\bfseries 13} (1964) 138--140}.

\bibitem{Wolfenstein:1964ks}
L.~Wolfenstein, ``{Violation of CP Invariance and the Possibility of Very Weak
  Interactions},'' \href{http://dx.doi.org/10.1103/PhysRevLett.13.562}{{\em
  Phys. Rev. Lett.} {\bfseries 13} (1964) 562--564}.

\bibitem{TheATLAScollaboration:2015mdt}
{ATLAS} Collaboration, ``{Search for resonances decaying to photon pairs in 3.2
  fb$^{-1}$ of $pp$ collisions at $\sqrt{s}$ = 13 TeV with the ATLAS
  detector},''  (12, 2015) \! \!\!\!, \url{https://cds.cern.ch/record/2114853}.

\bibitem{CMS:2015dxe}
{CMS} Collaboration, ``{Search for new physics in high mass diphoton events in
  proton-proton collisions at 13TeV},''  (2015) \! \!\!\!,
  \url{https://cds.cern.ch/record/2114808}.

\bibitem{Wolfenstein:1983yz}
L.~Wolfenstein, ``{Parametrization of the Kobayashi-Maskawa Matrix},''
  \href{http://dx.doi.org/10.1103/PhysRevLett.51.1945}{{\em Phys. Rev. Lett.}
  {\bfseries 51} (1983) 1945}.

\bibitem{Alonso:2013hga}
R.~Alonso, E.~E. Jenkins, A.~V. Manohar, and M.~Trott, ``{Renormalization Group
  Evolution of the Standard Model Dimension Six Operators III: Gauge Coupling
  Dependence and Phenomenology},''
  \href{http://dx.doi.org/10.1007/JHEP04(2014)159}{{\em JHEP} {\bfseries 04}
  (2014) 159}, \href{http://arxiv.org/abs/1312.2014}{{\ttfamily arXiv:1312.2014
  [hep-ph]}}.

\bibitem{Jenkins:2017jig}
E.~E. Jenkins, A.~V. Manohar, and P.~Stoffer, ``{Low-Energy Effective Field
  Theory below the Electroweak Scale: Operators and Matching},''
  \href{http://dx.doi.org/10.1007/JHEP03(2018)016}{{\em JHEP} {\bfseries 03}
  (2018) 016}, \href{http://arxiv.org/abs/1709.04486}{{\ttfamily
  arXiv:1709.04486 [hep-ph]}}.

\bibitem{E288:1977xhf}
{E288} Collaboration, S.~W. Herb {\em et~al.}, ``{Observation of a Dimuon
  Resonance at 9.5-GeV in 400-GeV Proton-Nucleus Collisions},''
  \href{http://dx.doi.org/10.1103/PhysRevLett.39.252}{{\em Phys. Rev. Lett.}
  {\bfseries 39} (1977) 252--255}.

\bibitem{Fernandez:1983az}
E.~Fernandez {\em et~al.}, ``{Lifetime of Particles Containing B Quarks},''
  \href{http://dx.doi.org/10.1103/PhysRevLett.51.1022}{{\em Phys. Rev. Lett.}
  {\bfseries 51} (1983) 1022}.

\bibitem{Lockyer:1983ev}
N.~Lockyer {\em et~al.}, ``{Measurement of the Lifetime of Bottom Hadrons},''
  \href{http://dx.doi.org/10.1103/PhysRevLett.51.1316}{{\em Phys. Rev. Lett.}
  {\bfseries 51} (1983) 1316}.

\bibitem{ARGUS:1987xtv}
{ARGUS} Collaboration, H.~Albrecht {\em et~al.}, ``{Observation of
  $B^0$--$\B0bar$ mixing},''
  \href{http://dx.doi.org/10.1016/0370-2693(87)91177-4}{{\em Phys. Lett. B}
  {\bfseries 192} (1987) 245--252}.

\bibitem{CLEO:1989gsk}
{CLEO} Collaboration, R.~Fulton {\em et~al.}, ``{Observation of B Meson
  Semileptonic Decays to Noncharmed Final States},''
  \href{http://dx.doi.org/10.1103/PhysRevLett.64.16}{{\em Phys. Rev. Lett.}
  {\bfseries 64} (1990) 16--20}.

\bibitem{ARGUS:1989eae}
{ARGUS} Collaboration, H.~Albrecht {\em et~al.}, ``{Observation of Semileptonic
  Charmless B Meson Decays},''
  \href{http://dx.doi.org/10.1016/0370-2693(90)91950-G}{{\em Phys. Lett. B}
  {\bfseries 234} (1990) 409--416}.

\bibitem{Vainshtein:1973md}
A.~I. Vainshtein and I.~B. Khriplovich, ``{Restrictions on masses of
  supercharged hadrons in the Weinberg model},'' {\em Pisma Zh. Eksp. Teor.
  Fiz.} {\bfseries 18} (1973) 141--145.

\bibitem{Gaillard:1974hs}
M.~K. Gaillard and B.~W. Lee, ``{Rare Decay Modes of the K-Mesons in Gauge
  Theories},'' \href{http://dx.doi.org/10.1103/PhysRevD.10.897}{{\em Phys. Rev.
  D} {\bfseries 10} (1974) 897}.

\bibitem{Glashow:1970gm}
S.~L. Glashow, J.~Iliopoulos, and L.~Maiani, ``{Weak Interactions with
  Lepton-Hadron Symmetry},''
  \href{http://dx.doi.org/10.1103/PhysRevD.2.1285}{{\em Phys. Rev. D}
  {\bfseries 2} (1970) 1285--1292}.

\bibitem{CDF:2006imy}
{CDF} Collaboration, A.~Abulencia {\em et~al.}, ``{Observation of $B^0_s -
  \bar{B}^0_s$ Oscillations},''
  \href{http://dx.doi.org/10.1103/PhysRevLett.97.242003}{{\em Phys. Rev. Lett.}
  {\bfseries 97} (2006) 242003},
  \href{http://arxiv.org/abs/hep-ex/0609040}{{\ttfamily arXiv:hep-ex/0609040}}.

\bibitem{LHCb:2021moh}
{LHCb} Collaboration, R.~Aaij {\em et~al.}, ``{Precise determination of the
  $B_s^0$\,--\,$\B0bar_s$ oscillation frequency},''
  \href{http://dx.doi.org/10.1038/s41567-021-01394-x}{{\em Nature Phys.}
  {\bfseries 18} no.~1, (2022) 1--5},
  \href{http://arxiv.org/abs/2104.04421}{{\ttfamily arXiv:2104.04421
  [hep-ex]}}.

\bibitem{Seng}
C.-Y. Seng, ``Talk at this symposium,''
\newblock 2023.
\newblock \url{https://conference-indico.kek.jp/event/195/}.

\bibitem{Grinstein:1987vj}
B.~Grinstein, R.~P. Springer, and M.~B. Wise, ``{Effective Hamiltonian for Weak
  Radiative B Meson Decay},''
  \href{http://dx.doi.org/10.1016/0370-2693(88)90868-4}{{\em Phys. Lett. B}
  {\bfseries 202} (1988) 138--144}.

\bibitem{Bertolini:1986th}
S.~Bertolini, F.~Borzumati, and A.~Masiero, ``{QCD Enhancement of Radiative b
  Decays},'' \href{http://dx.doi.org/10.1103/PhysRevLett.59.180}{{\em Phys.
  Rev. Lett.} {\bfseries 59} (1987) 180}.

\bibitem{Deshpande:1987nr}
N.~G. Deshpande, P.~Lo, J.~Trampetic, G.~Eilam, and P.~Singer, ``{$B \to K^*
  \gamma$ and the Top Quark Mass},''
  \href{http://dx.doi.org/10.1103/PhysRevLett.59.183}{{\em Phys. Rev. Lett.}
  {\bfseries 59} (1987) 183--185}.

\bibitem{Misiak:2015xwa}
M.~Misiak {\em et~al.}, ``{Updated NNLO QCD predictions for the weak radiative
  B-meson decays},''
  \href{http://dx.doi.org/10.1103/PhysRevLett.114.221801}{{\em Phys. Rev.
  Lett.} {\bfseries 114} no.~22, (2015) 221801},
  \href{http://arxiv.org/abs/1503.01789}{{\ttfamily arXiv:1503.01789
  [hep-ph]}}.

\bibitem{Bigi:1993ex}
I.~I.~Y. Bigi, M.~A. Shifman, N.~G. Uraltsev, and A.~I. Vainshtein, ``{On the
  motion of heavy quarks inside hadrons: Universal distributions and inclusive
  decays},'' \href{http://dx.doi.org/10.1142/S0217751X94000996}{{\em Int. J.
  Mod. Phys. A} {\bfseries 9} (1994) 2467--2504},
  \href{http://arxiv.org/abs/hep-ph/9312359}{{\ttfamily arXiv:hep-ph/9312359}}.

\bibitem{Neubert:1993um}
M.~Neubert, ``{Analysis of the photon spectrum in inclusive $B \to X_s \gamma$
  decays},'' \href{http://dx.doi.org/10.1103/PhysRevD.49.4623}{{\em Phys. Rev.
  D} {\bfseries 49} (1994) 4623--4633},
  \href{http://arxiv.org/abs/hep-ph/9312311}{{\ttfamily arXiv:hep-ph/9312311}}.

\bibitem{Kapustin:1995nr}
A.~Kapustin and Z.~Ligeti, ``{Moments of the photon spectrum in the inclusive
  $B \to X_s \gamma$ decay},''
  \href{http://dx.doi.org/10.1016/0370-2693(95)00762-A}{{\em Phys. Lett. B}
  {\bfseries 355} (1995) 318--324},
  \href{http://arxiv.org/abs/hep-ph/9506201}{{\ttfamily arXiv:hep-ph/9506201}}.

\bibitem{Bernlochner:2020jlt}
{SIMBA} Collaboration, F.~U. Bernlochner, H.~Lacker, Z.~Ligeti, I.~W. Stewart,
  F.~J. Tackmann, and K.~Tackmann, ``{Precision Global Determination of the $B
  \to X_s \gamma$ Decay Rate},''
  \href{http://dx.doi.org/10.1103/PhysRevLett.127.102001}{{\em Phys. Rev.
  Lett.} {\bfseries 127} no.~10, (2021) 102001},
  \href{http://arxiv.org/abs/2007.04320}{{\ttfamily arXiv:2007.04320
  [hep-ph]}}.

\bibitem{Isgur:1989vq}
N.~Isgur and M.~B. Wise, ``{Weak decays of heavy mesons in the static quark
  approximation},'' \href{http://dx.doi.org/10.1016/0370-2693(89)90566-2}{{\em
  Phys. Lett. B} {\bfseries 232} (1989) 113--117}.

\bibitem{Isgur:1990yhj}
N.~Isgur and M.~B. Wise, ``{Weak transition form factors between heavy
  mesons},'' \href{http://dx.doi.org/10.1016/0370-2693(90)91219-2}{{\em Phys.
  Lett. B} {\bfseries 237} (1990) 527--530}.

\bibitem{Georgi:1990um}
H.~Georgi, ``{An effective field theory for heavy quarks at low energies},''
  \href{http://dx.doi.org/10.1016/0370-2693(90)91128-X}{{\em Phys. Lett. B}
  {\bfseries 240} (1990) 447--450}.

\bibitem{Chay:1990da}
J.~Chay, H.~Georgi, and B.~Grinstein, ``{Lepton energy distributions in heavy
  meson decays from QCD},''
  \href{http://dx.doi.org/10.1016/0370-2693(90)90916-T}{{\em Phys. Lett. B}
  {\bfseries 247} (1990) 399--405}.

\bibitem{Isidori:2023unk}
G.~Isidori, Z.~Polonsky, and A.~Tinari, ``{Semi-inclusive $b\to s\bar\ell \ell$
  transitions at high $q^2$},''
  \href{http://dx.doi.org/10.1103/PhysRevD.108.093008}{{\em Phys. Rev. D}
  {\bfseries 108} no.~9, (2023) 093008},
  \href{http://arxiv.org/abs/2305.03076}{{\ttfamily arXiv:2305.03076
  [hep-ph]}}.

\bibitem{Belle:2021ymg}
{Belle} Collaboration, L.~Cao {\em et~al.}, ``{Measurement of Differential
  Branching Fractions of Inclusive ${B \to X_u \ell^+\, \nu_{\ell}}$ Decays},''
  \href{http://dx.doi.org/10.1103/PhysRevLett.127.261801}{{\em Phys. Rev.
  Lett.} {\bfseries 127} no.~26, (2021) 261801},
  \href{http://arxiv.org/abs/2107.13855}{{\ttfamily arXiv:2107.13855
  [hep-ex]}}.

\bibitem{Ligeti:2007sn}
Z.~Ligeti and F.~J. Tackmann, ``{Precise predictions for $B \to X_s l^+ l^-$ in
  the large $q^2$ region},''
  \href{http://dx.doi.org/10.1016/j.physletb.2007.07.070}{{\em Phys. Lett. B}
  {\bfseries 653} (2007) 404--410},
  \href{http://arxiv.org/abs/0707.1694}{{\ttfamily arXiv:0707.1694 [hep-ph]}}.

\bibitem{Luke:1990eg}
M.~E. Luke, ``{Effects of subleading operators in the heavy quark effective
  theory},'' \href{http://dx.doi.org/10.1016/0370-2693(90)90568-Q}{{\em Phys.
  Lett. B} {\bfseries 252} (1990) 447--455}.

\bibitem{El-Khadra}
A.~El-Khadra, ``Talk at this symposium,''
\newblock 2023.
\newblock \url{https://conference-indico.kek.jp/event/195/}.

\bibitem{FlavourLatticeAveragingGroupFLAG:2021npn}
{Flavour Lattice Averaging Group (FLAG)} Collaboration, Y.~Aoki {\em et~al.},
  ``{FLAG Review 2021},''
  \href{http://dx.doi.org/10.1140/epjc/s10052-022-10536-1}{{\em Eur. Phys. J.
  C} {\bfseries 82} no.~10, (2022) 869},
  \href{http://arxiv.org/abs/2111.09849}{{\ttfamily arXiv:2111.09849
  [hep-lat]}}.

\bibitem{HFLAV:2022esi}
{HFLAV} Collaboration, Y.~S. Amhis {\em et~al.}, ``{Averages of b-hadron,
  c-hadron, and \ensuremath{\tau}-lepton properties as of 2021},''
  \href{http://dx.doi.org/10.1103/PhysRevD.107.052008}{{\em Phys. Rev. D}
  {\bfseries 107} no.~5, (2023) 052008},
  \href{http://arxiv.org/abs/2206.07501}{{\ttfamily arXiv:2206.07501
  [hep-ex]}}.

\bibitem{Belle:2017rcc}
{Belle} Collaboration, A.~Abdesselam {\em et~al.}, ``{Precise determination of
  the CKM matrix element $\left| V_{cb}\right|$ with $\bar B^0 \to D^{*\,+} \,
  \ell^- \, \bar \nu_\ell$ decays with hadronic tagging at Belle},''
  \href{http://arxiv.org/abs/1702.01521}{{\ttfamily arXiv:1702.01521
  [hep-ex]}}.

\bibitem{Belle:2018ezy}
{Belle} Collaboration, E.~Waheed {\em et~al.}, ``{Measurement of the CKM matrix
  element $|V_{cb}|$ from $B^0\to D^{*-}\ell^ {+} \nu_\ell$ at Belle},''
  \href{http://dx.doi.org/10.1103/PhysRevD.100.052007}{{\em Phys. Rev. D}
  {\bfseries 100} no.~5, (2019) 052007},
  \href{http://arxiv.org/abs/1809.03290}{{\ttfamily arXiv:1809.03290
  [hep-ex]}}. [Erratum: Phys.Rev.D 103, 079901 (2021)].

\bibitem{Boyd:1995cf}
C.~G. Boyd, B.~Grinstein, and R.~F. Lebed, ``{Model independent extraction of
  $|V_{cb}|$ using dispersion relations},''
  \href{http://dx.doi.org/10.1016/0370-2693(95)00480-9}{{\em Phys. Lett. B}
  {\bfseries 353} (1995) 306--312},
  \href{http://arxiv.org/abs/hep-ph/9504235}{{\ttfamily arXiv:hep-ph/9504235}}.

\bibitem{Boyd:1995sq}
C.~G. Boyd, B.~Grinstein, and R.~F. Lebed, ``{Model independent determinations
  of $\bar B \to D l \bar\nu, D^* l \bar\nu$ form-factors},''
  \href{http://dx.doi.org/10.1016/0550-3213(95)00653-2}{{\em Nucl. Phys. B}
  {\bfseries 461} (1996) 493--511},
  \href{http://arxiv.org/abs/hep-ph/9508211}{{\ttfamily arXiv:hep-ph/9508211}}.

\bibitem{Boyd:1997kz}
C.~G. Boyd, B.~Grinstein, and R.~F. Lebed, ``{Precision corrections to
  dispersive bounds on form-factors},''
  \href{http://dx.doi.org/10.1103/PhysRevD.56.6895}{{\em Phys. Rev. D}
  {\bfseries 56} (1997) 6895--6911},
  \href{http://arxiv.org/abs/hep-ph/9705252}{{\ttfamily arXiv:hep-ph/9705252}}.

\bibitem{Neubert:1992wq}
M.~Neubert, Z.~Ligeti, and Y.~Nir, ``{QCD sum rule analysis of the subleading
  Isgur-Wise form-factor $\chi_2(v\cdot v')$},''
  \href{http://dx.doi.org/10.1016/0370-2693(93)90728-Z}{{\em Phys. Lett. B}
  {\bfseries 301} (1993) 101--107},
  \href{http://arxiv.org/abs/hep-ph/9209271}{{\ttfamily arXiv:hep-ph/9209271}}.

\bibitem{Neubert:1992pn}
M.~Neubert, Z.~Ligeti, and Y.~Nir, ``{The Subleading Isgur-Wise form-factor
  $\chi_3(v\cdot v')$ to order $\alpha_s$ in QCD sum rules},''
  \href{http://dx.doi.org/10.1103/PhysRevD.47.5060}{{\em Phys. Rev. D}
  {\bfseries 47} (1993) 5060--5066},
  \href{http://arxiv.org/abs/hep-ph/9212266}{{\ttfamily arXiv:hep-ph/9212266}}.

\bibitem{Ligeti:1993hw}
Z.~Ligeti, Y.~Nir, and M.~Neubert, ``{The Subleading Isgur-Wise form-factor
  $\xi_3(v\cdot v')$ and its implications for the decays $\bar B \to D^* \ell
  \bar\nu$},'' \href{http://dx.doi.org/10.1103/PhysRevD.49.1302}{{\em Phys.
  Rev. D} {\bfseries 49} (1994) 1302--1309},
  \href{http://arxiv.org/abs/hep-ph/9305304}{{\ttfamily arXiv:hep-ph/9305304}}.

\bibitem{Caprini:1997mu}
I.~Caprini, L.~Lellouch, and M.~Neubert, ``{Dispersive bounds on the shape of
  anti-B ---\ensuremath{>} D(*) lepton anti-neutrino form-factors},''
  \href{http://dx.doi.org/10.1016/S0550-3213(98)00350-2}{{\em Nucl. Phys. B}
  {\bfseries 530} (1998) 153--181},
  \href{http://arxiv.org/abs/hep-ph/9712417}{{\ttfamily arXiv:hep-ph/9712417}}.

\bibitem{Bordone:2019vic}
M.~Bordone, M.~Jung, and D.~van Dyk, ``{Theory determination of $\bar{B}\to
  D^{(*)}\ell^-\bar\nu$ form factors at $\mathcal{O}(1/m_c^2)$},''
  \href{http://dx.doi.org/10.1140/epjc/s10052-020-7616-4}{{\em Eur. Phys. J. C}
  {\bfseries 80} no.~2, (2020) 74},
  \href{http://arxiv.org/abs/1908.09398}{{\ttfamily arXiv:1908.09398
  [hep-ph]}}.

\bibitem{Bernlochner:2022ywh}
F.~U. Bernlochner, Z.~Ligeti, M.~Papucci, M.~T. Prim, D.~J. Robinson, and
  C.~Xiong, ``{Constrained second-order power corrections in HQET:
  $R(D^{(*)})$, $|V_{cb}|$, and new physics},''
  \href{http://dx.doi.org/10.1103/PhysRevD.106.096015}{{\em Phys. Rev. D}
  {\bfseries 106} no.~9, (2022) 096015},
  \href{http://arxiv.org/abs/2206.11281}{{\ttfamily arXiv:2206.11281
  [hep-ph]}}.

\bibitem{Bernlochner:2019ldg}
F.~U. Bernlochner, Z.~Ligeti, and D.~J. Robinson, ``{N = 5, 6, 7, 8: Nested
  hypothesis tests and truncation dependence of $|V_{cb}|$},''
  \href{http://dx.doi.org/10.1103/PhysRevD.100.013005}{{\em Phys. Rev. D}
  {\bfseries 100} no.~1, (2019) 013005},
  \href{http://arxiv.org/abs/1902.09553}{{\ttfamily arXiv:1902.09553
  [hep-ph]}}.

\bibitem{Gambino:2019sif}
P.~Gambino, M.~Jung, and S.~Schacht, ``{The $V_{cb}$ puzzle: An update},''
  \href{http://dx.doi.org/10.1016/j.physletb.2019.06.039}{{\em Phys. Lett. B}
  {\bfseries 795} (2019) 386--390},
  \href{http://arxiv.org/abs/1905.08209}{{\ttfamily arXiv:1905.08209
  [hep-ph]}}.

\bibitem{Simons:2023wfg}
D.~Simons, E.~Gustafson, and Y.~Meurice, ``{Self-consistent optimization of the
  $z$-Expansion for $B$ meson decays},''
  \href{http://arxiv.org/abs/2304.13045}{{\ttfamily arXiv:2304.13045
  [hep-ph]}}.

\bibitem{Belle:2023bwv}
{Belle} Collaboration, M.~T. Prim {\em et~al.}, ``{Measurement of differential
  distributions of $B\to D^* \ell \nu_\ell$ and implications on $|V_{cb}|$},''
  \href{http://dx.doi.org/10.1103/PhysRevD.108.012002}{{\em Phys. Rev. D}
  {\bfseries 108} no.~1, (2023) 012002},
  \href{http://arxiv.org/abs/2301.07529}{{\ttfamily arXiv:2301.07529
  [hep-ex]}}.

\bibitem{Belle-II:2023okj}
{Belle~II} Collaboration, I.~Adachi {\em et~al.}, ``{Determination of
  $|V_{cb}|$ using $\B0bar \to D^{*+} \ell^- \bar\nu$ decays with Belle II},''
  \href{http://dx.doi.org/10.1103/PhysRevD.108.092013}{{\em Phys. Rev. D}
  {\bfseries 108} no.~9, (2023) 092013},
  \href{http://arxiv.org/abs/2310.01170}{{\ttfamily arXiv:2310.01170
  [hep-ex]}}.

\bibitem{ZLoct23}
Z.~Ligeti, ``{Talk at the Workshop on precision $|V_{cb}|$ measurements at
  Belle II},'' 2023.
\newblock
  \url{https://indico.belle2.org/event/9402/contributions/63857/attachments/25094/37246/kek23.pdf}.

\bibitem{Bauer:2004ve}
C.~W. Bauer, Z.~Ligeti, M.~Luke, A.~V. Manohar, and M.~Trott, ``{Global
  analysis of inclusive B decays},''
  \href{http://dx.doi.org/10.1103/PhysRevD.70.094017}{{\em Phys. Rev. D}
  {\bfseries 70} (2004) 094017},
  \href{http://arxiv.org/abs/hep-ph/0408002}{{\ttfamily arXiv:hep-ph/0408002}}.

\bibitem{Bauer:2002sh}
C.~W. Bauer, Z.~Ligeti, M.~Luke, and A.~V. Manohar, ``{B decay shape variables
  and the precision determination of $|V_{cb}|$ and $m_b$},''
  \href{http://dx.doi.org/10.1103/PhysRevD.67.054012}{{\em Phys. Rev. D}
  {\bfseries 67} (2003) 054012},
  \href{http://arxiv.org/abs/hep-ph/0210027}{{\ttfamily arXiv:hep-ph/0210027}}.

\bibitem{Alberti:2014yda}
A.~Alberti, P.~Gambino, K.~J. Healey, and S.~Nandi, ``{Precision Determination
  of the Cabibbo-Kobayashi-Maskawa Element $V_{cb}$},''
  \href{http://dx.doi.org/10.1103/PhysRevLett.114.061802}{{\em Phys. Rev.
  Lett.} {\bfseries 114} no.~6, (2015) 061802},
  \href{http://arxiv.org/abs/1411.6560}{{\ttfamily arXiv:1411.6560 [hep-ph]}}.

\bibitem{Gambino:2004qm}
P.~Gambino and N.~Uraltsev, ``{Moments of semileptonic B decay distributions in
  the 1/m(b) expansion},''
  \href{http://dx.doi.org/10.1140/epjc/s2004-01671-2}{{\em Eur. Phys. J. C}
  {\bfseries 34} (2004) 181--189},
  \href{http://arxiv.org/abs/hep-ph/0401063}{{\ttfamily arXiv:hep-ph/0401063}}.

\bibitem{Fael:2020njb}
M.~Fael, K.~Sch\"onwald, and M.~Steinhauser, ``{Relation between the
  $\overline{\mathrm{MS}}$ and the kinetic mass of heavy quarks},''
  \href{http://dx.doi.org/10.1103/PhysRevD.103.014005}{{\em Phys. Rev. D}
  {\bfseries 103} no.~1, (2021) 014005},
  \href{http://arxiv.org/abs/2011.11655}{{\ttfamily arXiv:2011.11655
  [hep-ph]}}.

\bibitem{Fael:2020tow}
M.~Fael, K.~Sch\"onwald, and M.~Steinhauser, ``{Third order corrections to the
  semileptonic $b\to c$ and the muon decays},''
  \href{http://dx.doi.org/10.1103/PhysRevD.104.016003}{{\em Phys. Rev. D}
  {\bfseries 104} no.~1, (2021) 016003},
  \href{http://arxiv.org/abs/2011.13654}{{\ttfamily arXiv:2011.13654
  [hep-ph]}}.

\bibitem{Bordone:2021oof}
M.~Bordone, B.~Capdevila, and P.~Gambino, ``{Three loop calculations and
  inclusive $V_{cb}$},''
  \href{http://dx.doi.org/10.1016/j.physletb.2021.136679}{{\em Phys. Lett. B}
  {\bfseries 822} (2021) 136679},
  \href{http://arxiv.org/abs/2107.00604}{{\ttfamily arXiv:2107.00604
  [hep-ph]}}.

\bibitem{Mannel:2021zzr}
T.~Mannel, D.~Moreno, and A.~A. Pivovarov, ``{NLO QCD corrections to inclusive
  $b \rightarrow c \ell \bar{\nu}$ decay spectra up to $1/m_Q^3$},''
  \href{http://dx.doi.org/10.1103/PhysRevD.105.054033}{{\em Phys. Rev. D}
  {\bfseries 105} no.~5, (2022) 054033},
  \href{http://arxiv.org/abs/2112.03875}{{\ttfamily arXiv:2112.03875
  [hep-ph]}}.

\bibitem{Barone:2023tbl}
A.~Barone, S.~Hashimoto, A.~J\"uttner, T.~Kaneko, and R.~Kellermann,
  ``{Approaches to inclusive semileptonic B$_{(s)}$-meson decays from Lattice
  QCD},'' \href{http://dx.doi.org/10.1007/JHEP07(2023)145}{{\em JHEP}
  {\bfseries 07} (2023) 145}, \href{http://arxiv.org/abs/2305.14092}{{\ttfamily
  arXiv:2305.14092 [hep-lat]}}. And references therein.

\bibitem{LHCb:2020cyw}
{LHCb} Collaboration, R.~Aaij {\em et~al.}, ``{Measurement of $|V_{cb}|$ with
  $B_s^0 \to D_s^{(*)-} \mu^+ \nu_{\mu}$ decays},''
  \href{http://dx.doi.org/10.1103/PhysRevD.101.072004}{{\em Phys. Rev. D}
  {\bfseries 101} no.~7, (2020) 072004},
  \href{http://arxiv.org/abs/2001.03225}{{\ttfamily arXiv:2001.03225
  [hep-ex]}}.

\bibitem{Bernlochner:2017jka}
F.~U. Bernlochner, Z.~Ligeti, M.~Papucci, and D.~J. Robinson, ``{Combined
  analysis of semileptonic $B$ decays to $D$ and $D^*$: $R(D^{(*)})$,
  $|V_{cb}|$, and new physics},''
  \href{http://dx.doi.org/10.1103/PhysRevD.95.115008}{{\em Phys. Rev. D}
  {\bfseries 95} no.~11, (2017) 115008},
  \href{http://arxiv.org/abs/1703.05330}{{\ttfamily arXiv:1703.05330
  [hep-ph]}}. [Erratum: Phys.Rev.D 97, 059902 (2018)].

\bibitem{Charles:2004jd}
J.~Charles, A.~Hocker, H.~Lacker, S.~Laplace, F.~R. Le~Diberder, J.~Malcles,
  J.~Ocariz, M.~Pivk, and L.~Roos, ``{CP violation and the CKM matrix:
  Assessing the impact of the asymmetric $B$ factories},''
  \href{http://dx.doi.org/10.1140/epjc/s2005-02169-1}{{\em Eur. Phys. J. C}
  {\bfseries 41} no.~1, (2005) 1--131},
  \href{http://arxiv.org/abs/hep-ph/0406184}{{\ttfamily arXiv:hep-ph/0406184}}.
  And updates at \url{http://ckmfitter.in2p3.fr/}.

\bibitem{Bediaga:2018lhg}
{LHCb} Collaboration, R.~Aaij {\em et~al.}, ``{Physics case for an LHCb Upgrade
  II -- Opportunities in flavour physics, and beyond, in the HL-LHC era},''
  \href{http://arxiv.org/abs/1808.08865}{{\ttfamily arXiv:1808.08865
  [hep-ex]}}.

\bibitem{Cerri:2018ypt}
A.~Cerri {\em et~al.}, ``{Report from Working Group 4}: {Opportunities in
  Flavour Physics at the HL-LHC and HE-LHC},''
  \href{http://dx.doi.org/10.23731/CYRM-2019-007.867}{{\em CERN Yellow Rep.
  Monogr.} {\bfseries 7} (2019) 867--1158},
  \href{http://arxiv.org/abs/1812.07638}{{\ttfamily arXiv:1812.07638
  [hep-ph]}}.

\bibitem{Belle-II:2018jsg}
{Belle~II} Collaboration, W.~Altmannshofer {\em et~al.}, ``{The Belle II
  Physics Book},'' \href{http://dx.doi.org/10.1093/ptep/ptz106}{{\em PTEP}
  {\bfseries 2019} no.~12, (2019) 123C01},
  \href{http://arxiv.org/abs/1808.10567}{{\ttfamily arXiv:1808.10567
  [hep-ex]}}. [Erratum: PTEP 2020, 029201 (2020)].

\bibitem{BESIII:2022mxl}
{BESIII} Collaboration, H.~B. Li {\em et~al.}, ``{Physics in the $\tau$-charm
  Region at BESIII},'' in {\em {2022 Snowmass Summer Study}}.
\newblock 4, 2022.
\newblock \href{http://arxiv.org/abs/2204.08943}{{\ttfamily arXiv:2204.08943
  [hep-ex]}}.

\bibitem{FCC:2018byv}
{FCC} Collaboration, A.~Abada {\em et~al.}, ``{FCC Physics Opportunities}:
  {Future Circular Collider Conceptual Design Report Volume 1},''
  \href{http://dx.doi.org/10.1140/epjc/s10052-019-6904-3}{{\em Eur. Phys. J. C}
  {\bfseries 79} no.~6, (2019) 474}.

\bibitem{Grossman:2021xfq}
Y.~Grossman and Z.~Ligeti, ``{Theoretical challenges for flavor physics},''
  \href{http://dx.doi.org/10.1140/epjp/s13360-021-01845-7}{{\em Eur. Phys. J.
  Plus} {\bfseries 136} no.~9, (2021) 912},
  \href{http://arxiv.org/abs/2106.12168}{{\ttfamily arXiv:2106.12168
  [hep-ph]}}.

\bibitem{Azzurri}
P.~Azzurri, ``Flavor tagging in $\uppercase{W}$ decays.''. Talk at the 4th FCC
  Physics and Experiments Workshop, Nov.\ 2020,
  \url{https://indico.cern.ch/event/932973/contributions/4059403/attachments/2140815/3607142/azzurriFCCeeWHF.pdf}.

\bibitem{Bauer:2000xf}
C.~W. Bauer, Z.~Ligeti, and M.~E. Luke, ``{A Model independent determination of
  $|V_{ub}|$},'' \href{http://dx.doi.org/10.1016/S0370-2693(00)00318-X}{{\em
  Phys. Lett. B} {\bfseries 479} (2000) 395--401},
  \href{http://arxiv.org/abs/hep-ph/0002161}{{\ttfamily arXiv:hep-ph/0002161}}.

\bibitem{Kerstin}
K.~Tackmann, ``Extracting $|\uppercase{V}_{ub}|$ and \mbox{$B\to X_s\gamma$}
  from global fits.''. Talk at the 4th SuperB Collaboration Meeting, La Biodola
  (Isola d'Elba) Italy, Jun.\ 2012,
  \url{https://agenda.infn.it/event/4880/contributions/55588/attachments/39553/46670/tackmann-superb-elba-31may12.pdf}.

\bibitem{Charles:2020dfl}
J.~Charles, S.~Descotes-Genon, Z.~Ligeti, S.~Monteil, M.~Papucci, K.~Trabelsi,
  and L.~Vale~Silva, ``{New physics in $B$ meson mixing: future sensitivity and
  limitations},'' \href{http://dx.doi.org/10.1103/PhysRevD.102.056023}{{\em
  Phys. Rev. D} {\bfseries 102} no.~5, (2020) 056023},
  \href{http://arxiv.org/abs/2006.04824}{{\ttfamily arXiv:2006.04824
  [hep-ph]}}.

\bibitem{LHCb:2022ine}
{LHCb} Collaboration, ``{Future physics potential of LHCb},''.
  \url{https://inspirehep.net/literature/2071541}.

\bibitem{Forti:2022mti}
{Belle~II} Collaboration, F.~Forti, ``{Snowmass Whitepaper: The Belle II
  Detector Upgrade Program},'' in {\em {Snowmass 2021}}.
\newblock 3, 2022.
\newblock \href{http://arxiv.org/abs/2203.11349}{{\ttfamily arXiv:2203.11349
  [hep-ex]}}.

\bibitem{Sanda}
A.~I. Sanda, ``{Talk at the 5th Workshops on Higher Luminosity $B$ Factory},''
  Sept., 2003.
\newblock \url{https://belle.kek.jp/superb/workshop/2003/HL05/}.

\end{thebibliography}\endgroup



\end{document}